\documentclass[
 reprint,
 amsmath,amssymb,
 superscriptaddress,
]{revtex4-1}

\usepackage[all,defaultlines=3]{nowidow} 
\usepackage[section]{placeins} 
\usepackage{color} 
\usepackage{lineno} 
\usepackage[utf8]{inputenc} 
\usepackage[caption = false]{subfig} 
\usepackage{graphicx}
\usepackage{float}
\usepackage{dcolumn}
\usepackage{changes} 
\definechangesauthor[name=Renaud, color=blue]{R}

\usepackage{natbib} 

\usepackage{bm}
\usepackage{textcomp}
\usepackage{upgreek}
\usepackage{csquotes} 
\usepackage{cancel} 

\usepackage{hyperref} 
\hypersetup{colorlinks=true, linkcolor=blue, filecolor=magenta, urlcolor=blue}  

\begin{document}

\title{Stable and unstable capillary fingering\\in porous media with a gradient in grains size}

\author{Tom Vincent-Dospital}
\email{tom.vincent-dospital@fys.uio.no}
\affiliation{PoreLab, The Njord Centre, Department of Physics, University of Oslo, P.O. Box 1048 Blindern, N-0316 Oslo, Norway}

\author{Marcel Moura}
\email{marcel.moura@fys.uio.no}
\affiliation{PoreLab, The Njord Centre, Department of Physics, University of Oslo, P.O. Box 1048 Blindern, N-0316 Oslo, Norway}

\author{Renaud Toussaint}
\affiliation{Université de Strasbourg, CNRS, Institut Terre \& Environnement de Strasbourg, UMR 7063, Strasbourg, France}
\affiliation{PoreLab, The Njord Centre, Department of Physics, University of Oslo, P.O. Box 1048 Blindern, N-0316 Oslo, Norway}

\author{Knut J\o rgen M\aa l\o y}
\email{maloy@fys.uio.no}
\affiliation{PoreLab, The Njord Centre, Department of Physics, University of Oslo, P.O. Box 1048 Blindern, N-0316 Oslo, Norway}\affiliation{PoreLab, Department of Geoscience and Petroleum, Norwegian University of Science and Technology NTNU, 7031 Trondheim, Norway}

\date{\today}

\begin{abstract}
We present a theoretical and experimental investigation of slow drainage in porous media with a gradient in the grains size (and hence in the typical pores' throats), in an external gravitational field. We mathematically show that such structural gradient and external force have a similar effect on the obtained drainage patterns, when they stabilise the invasion front. With the help of a newly introduced experimental set-up, based on the 3D-print of transparent porous matrices, we illustrate this equivalence, and extend it to the case where the front is unstable. We also present some invasion-percolation simulations of the same phenomena, which are inline with our theoretical and experimental results. In particular, we show that the width of stable drainage fronts mainly scales with the spatial gradient of the average pore invasion threshold and with the local distribution of this (disordered) threshold. The scaling exponent results from percolation theory and is $-0.57$ for 2D systems. Overall, we propose a unifying theory for the up-scaling of dual fluid flows in most classical scenarii.
\end{abstract}


\maketitle


\section{Introduction}
Two-phase flow in porous media is important in a wide range of sectors. It remains to be fully understood by fundamental research, and holds applications in activities as diverse as food processing, management of underground water resources, sequestration of greenhouse gases or crude oil recovery. It is also crucial in everyday phenomena such as plant watering or making a cup of coffee. Due to its broad impact but also to its complexity, it is a multidisciplinary subject that has been long investigated by physicists, geoscientists, hydrologists, chemists, biologists, and engineers. 
When one fluid displaces the other, the resulting structures in their distribution present various shapes and complexity\,\cite{lenormand1988, lenormand1989, zhao2016, maloy1985, chen1985}, that can be compact, ramified and fractal\,\cite{mandelbrot1982, feder1988}. These structures are controlled by the forces which actually drive the flow. These forces can for instance be viscous\,\cite{chen1985, maloy1985, weitz1987, lenormand1988, lovoll2004, Maloy2021}, capillary\,\cite{lenormand1988, lenormand1985, lovoll2004, Maloy2021, aval_imb}, gravitational\,\cite{Wilkinson1984, birovljev91, Frette92, auradou1999, meheust2002, Pride2022} or a combination of them \cite{toussaint2012}, and may depend on physical parameters such as the wetting properties of the solid-fluids system\,\cite{zhao2016} or changes in the local geometry of the porous medium\,\cite{rabbani2018,lu2019}.\\
In this work, we study the effect of the latter, that is, of the geometry of the solid matrix. In particular, we run slow drainage experiments (i.e., the replacement of a wetting fluid by a non-wetting one at a rate where viscous forces are negligible) in 2D porous models that present a spatial gradient in their grains size.
This gradient is obtained by a progressive change in the size of the pore throats of the models along the main flow direction. Note that, in the case where viscous forces would not be negligible, such a gradient could actually also be seen, in a Darcy's view, as a permeability gradient. Such a configuration has already been considered\,\cite{rabbani2018,lu2019} using regular porous matrices, but we here use models that conserve a - controlled - random disorder, better simulating actual natural systems. Examples of such natural systems are graded geological bedding\,\cite{Graded_bedding}, for instance formed in rivers of varying stream intensity or by turbidite deposition, which are notably of interest to hydrologists and petroleum geoscientists. In our experiments, the control in the disorder is achieved by the 3D printing of our porous matrices in a transparent material, so that the drainage and the fluids' distributions can be visually monitored. We show that the gradient in grains size has a similar effect on the fluids flow than that of gravity when tilting the model (or, alternatively, any other external field). In particular, depending on the flow direction relative to this gradient, the drainage can be stabilised or destabilised. We also provide the theoretical explanation for this equivalence, generalising the recently reintroduced fluctuation number\,\cite{Maloy2021} to describe the effect of the noise on the width of the invasion front. Finally, we present invasion-percolation simulations reproducing the experimental results and the predicted scaling law between the width of the invasion front and the gradient in pore invasion pressure, the local disorder in this invasion pressure, and the typical pore size of the matrix.

\section{\label{sec:set-up} Experimental Technique}

\begin{figure*}
\centering
\includegraphics[width=1.0\linewidth]{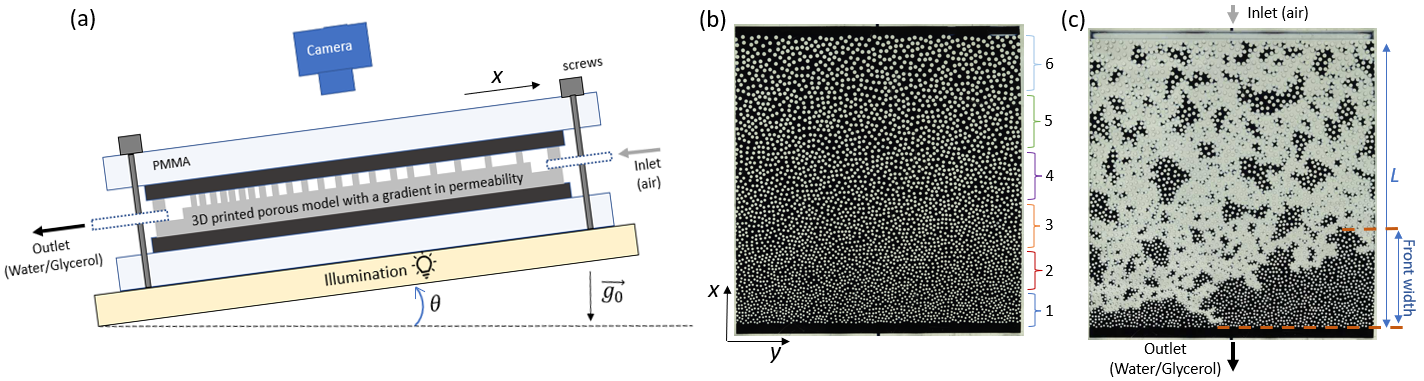}
\caption{(a): Schematic cross section of the experimental set-up. Two transparent PMMA plates ($3$\,cm thick) are screwed together to close the 3D printed porous model with a gradient in grain and pores' throat sizes. Soft polymer layers insure good contacts between the PMMA plates and the model. The model is illuminated from below and pictures are taken from above. (b-c) Top view pictures. A non-wetting fluid (air) invades another wetting fluid (water/glycerol) in the square porous model of size $L=140$\,mm and coordinate system ($x$,$y$). A pump is connected to the outlet and the inlet is open to air. The model can be tilted, leading to a capillary pressure gradient $G=\Delta\rho\,g_0\,\text{sin}(\theta)$ in the $x$ direction, where $g_0=9.82$\,m\,s\textsuperscript{-2} and $\Delta\rho\sim1200$\,kg\,m\textsuperscript{-3} is the difference of density between the wetting and non-wetting fluids. The full-extent width of the invasion front highlighted in (c) and, to better characterise it, we will denote $\eta$ the standard deviation of the front position along $x$. The model areas marked 1 to 6 correspond to the distributions shown in Fig.\,\ref{dist-shift}, and their size scales the same way as the pore size in the model.}
\label{syst} 
\end{figure*}

\begin{figure}
\centering
\includegraphics[width=1.0\linewidth]{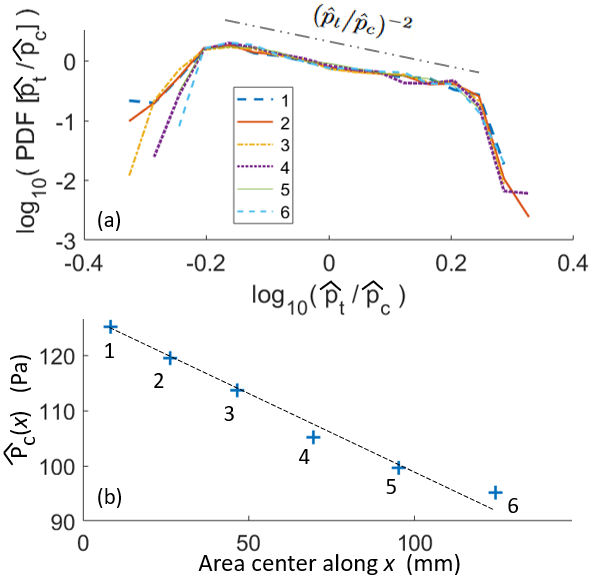}
\caption{(a): Probability density function $\tilde{N}$ for $\hat{p}_t(x,y)/\hat{p}_c$ in the six areas (denoted 1 to 6) defined in Fig.\,\ref{syst} on the 3D printed model. This distribution is conserved in our model and, around $\hat{p}_t=p_c$, it scales approximately as $(\hat{p}_t/\hat{p}_c)^{-2}$, as highlighted by the dash-dotted line. Note: the average pore connectivity in our model being 3, we defined the percolation pressure $\hat{p}_c$ such that $\int_{0}^{\hat{p}_c} N(\hat{p}_t) \,\mathrm{d}\hat{p}_t \approx 0.65$ (e.g., see\,\cite{moura2015,perco_theo}). (b): Evolution of the critical capillary pressure $\hat{p}_c$ along the model (crosses), arising from its intrinsic structure. The straight line shows a linear fit, which is a first order approximation of $\hat{p}_c$ (the engineered gradient in the model was actually in grain size and not in percolation pressure). To this gradient, one can compute an equivalent model tilt (see Fig.\,\ref{syst}): $\theta_\text{eq}=\text{asin}(-\frac{\partial \hat{p}_c(x)}{\partial x}/\Delta\rho\,g_0)$. Here, $\theta_\text{eq}\sim1.35$\,degrees.}
\label{dist-shift} 
\end{figure}

To generate our porous models, we have used a Formlabs Form 3L printer\,\cite{Formlabs} which employs a stereolithography 3D printing technology to produce models in a transparent plastic material (Formlabs Clear Resin). This technique, based on the laser polymerisation of the resin, allows us to control the geometry of the porous network and in particular to fine-tune its grain and pores' throats distribution, for example by introducing a gradient in the pore sizes. We have made quasi two-dimensional models where cylinders are distributed in a monolayer using a Random Sequential Adsorption (RSA) algorithm\,\cite{hinrichsen1986}. The spatial resolution of the printed models is about $0.1\,$mm. The RSA parameters are the minimum distance between the cylinders, the cylinders' diameter, the cylinders' height and the dimensions of the desired porous system. In our case, the RSA parameters vary between the model's inlet and outlet. The total size of the model is $140\times140\,$mm\textsuperscript{2}, with the cylinder diameter $d$ varying linearly from $1\,$mm to $2\,$mm and the minimum cylinder's separation varying from $0.4\,$mm to $0.8\,$mm, in order to preserve geometrical similarity along the model's length. We thus design a model with a grain (cylinder) size gradient $\lambda\sim0.007\,$mm mm\textsuperscript{-1}. The height of the cylinder was chosen to be $2\,$mm. The obtained model and experimental set-up is shown in Fig.\,\ref{syst}.\\
In Fig.\,\ref{syst}, one can also see the flow cell which was constructed around the prints, in a way that optimises the visualisation of the pores, seen from a top-down view. The 3D printed model is inserted between two layers of a soft polymer ($5$\,mm thick) whose main role is to efficiently seal the top of each cylinder. Around these soft layers two thick PMMA plates of width $3$\,cm are screwed together to confine the flow. On the side, the model is closed by 3D printed walls. The tubular inlet and the outlet of the model feed large channels along the whole model width so that the boundary conditions are the same along this direction. Each layer is transparent and the flow cell lies on a white light box to allow a good quality imaging with a reflex camera. The whole set-up can be tilted by an angle $\theta$ for the flow to occur in a chosen effective gravity field $g_0 \sin(\theta)$.\\
In our experiments, the invading fluid is air and the defending one in a mixture of $20$\% water and $80$\% glycerol, where the percentages relate to the total mass. In this mixture a nigrosin dye has been added ($4$\,grams per litre of water) to create an imaging contrast between the air and the liquid.  With a syringe pump, the wetting liquid mixture is withdrawn at a constant flow rate from one of the two models' ends while the air, connected to the atmospheric pressure of the laboratory, invades the model. We used a $0.3$\,ml/hour flow rate, ensuring a capillary only drainage (i.e., negligible viscous effect) with a small capillary number $C_a=\mu V /\gamma$ that is about\,$10^{-8}$, where $V$ is the typical flow velocity, $\mu$ is the water-glycerol's viscosity and $\gamma$ is the surface tension at the fluids' interface.

\section{\label{sec:theory} Theory - Gradient in pore throats and external fields}

The requirement for invasion into one pore neck by the non-wetting fluid is that the capillary pressure $p$ between the two fluids overcomes the capillary threshold value $\hat{p}_t$ of this pore neck 
\begin{equation}
    p(x,y)>\hat{p}_t(x,y) \; ,
\end{equation}
where $x$ and $y$ are the 2D coordinates for the pore neck position. In the case of our 3D printed model, the distribution in capillary pressure along the model is shown in Fig.\,\ref{dist-shift}. It is obtained from an approximation of Young-Laplace's equation as $\hat{p}_t(x,y)=\gamma \cos(\phi)(1/r+1/h)$, where $r$ is the pore throat, $h$ is the cylinders' height and $\phi$ is the contact angle - measured within the defending phase - for the solid matrix and the two fluids at play.\\
Assume that we have an external field that change the capillary pressure linearly in the $x$ direction. One such field is the gravitational field (e.g., controlled by $\theta$ in Fig.\,\ref{syst}). Another one, if the flow is fast enough, is the viscous pressure drop inside the fluid being withdrawn. In the case discussed in the present manuscript, the flow rate is slow enough for this viscous effect to be negligible compared to that of the other forces at stake. We will however keep a general formalism so they may be included. If we write the gradient in capillary pressure from the external fields as $G$, this capillary pressure at a position $(x,y)$ is
\begin{equation}
    p(x,y)=p(x_0)+G \cdot(x-x_0) \; , 
\end{equation}
where $p(x_0)$ is the capillary pressure at an arbitrary position $x_0$. The condition for invasion can then be written as 
\begin{equation}
    p(x_0)>\hat{p}_t(x,y)-G\cdot (x-x_0)=p_t(x,y) \; ,
\end{equation}
where we have introduced the  modified thresholds $p_t$. 
The system is now mapped onto a system without fields but where the thresholds are modified with the fields as a linear terms in $x$. The capillary pressure $p=p(x_0)$ for this equivalent system is constant over the model. We now consider this modified system.  
The mapping between the occupation probability in percolation $f$ and the capillary pressure $p$ is given\,\cite{perco_theo} by 
\begin{equation}
  f(x)-f_c=\int^{p}_{p_c(x)} N(p_t,x)\,\mathrm{d}p_t 
  \label{int2}
\end{equation}
In this expression, $f_c$ is the critical occupation probability, $p_c$ is the critical percolation pressure and $N$ is the distribution in capillary pressure threshold. This distribution is a function of $x$ due to the overall gradient in pore neck and to the external field.\\
In the case where the invasion front is stable, a position $x_1$ exists were the capillary pressure is equal to the critical capillary pressure such that
 \begin{equation}
 p=p_c(x_1) \; .
 \label{peqpc}
 \end{equation}
We can also Taylor expand $N(p_t,x)$ around the critical $p_c(x_1)$ in Eq.\,(\ref{int2}) keeping only the lowest order in $p-p_c(x_1)$ such that
\begin{equation}
     f(x)-f_c=N(p_c(x_1),x)[p_c(x_1)-p_c(x)] \; .
\end{equation}
Expanding $p_c(x)$ to first order in $x-x_1$
\begin{equation}
  p_c(x)=p_c(x_1) + \left. \frac{\partial p_c(x)}{\partial x} \right|_{x_1} (x-x_1) \; ,  
\end{equation}
we get 
\begin{equation}
   f(x)-f_c=-N(p_c(x_1),x) \left. \frac{\partial p_c(x)}{\partial x}\right|_{x_1} (x-x_1) \; .  
\end{equation}
Now, $\partial p_c(x)/\partial x$ will contain one term from the gradient in the critical capillary pressure $\partial \hat{p}_c(x)/\partial x$ of the porous medium plus the term $G$ which is due to the external linear field. This gives 
 \begin{equation}
     f(x)-f_c=aN(p_c(x_1),x)\left(\left. G - \frac{\partial \hat{p}_c(x)}{\partial x} \right|_{x_1}\right)\frac{x-x_1}{a} \; .
   \label{eq:subfin}
 \end{equation}
Here, we have introduced $a$ as the typical length of a pore. We now choose $x$ such that $\eta=|x_1-x|$, where $\eta$ is the width of the front. We here write $\eta=x_1-x>0$, which corresponds to an invasion flow that progresses against the $x$ direction if $f(x)<f_c$. The opposite convention for $x$ could have, of course, also been chosen. Furthermore, we use Sapoval's assumption\,\cite{sapova1985}, that $\eta$ scales in the same way as the correlation length $\xi$ in percolation, $(\xi/a) \propto |f-f_c|^{-\nu}$, where $\nu$ is a critical exponent typically equals to $4/3$ for 2D systems\,\cite{perco_theo}. We obtain $(\eta/a) \propto (f_c-f)^{-\nu}$ and then, with Eq.\,(\ref{eq:subfin}), we find
\begin{equation}
  \eta/a \propto F^{\frac{-\nu}{1+\nu}} \; ,
  \label{eta1}
\end{equation}
where the exponent $\beta=\nu/(1+\nu)$ is about $0.57$. For a 3D system, $\beta$ would be approximately $0.47$, with $\nu\sim0.88$\,\cite{perco_theo}. We call the quantity $F$ the fluctuation number, which writes as
\begin{equation}
    F(x)=a(x)N\left(p_c(x_1),x\right)\left(\left. G - \frac{\partial \hat{p}_c(x)}{\partial x} \right|_{x_1}\right) \; .
    \label{eqF}
\end{equation}
$F$ is a dimensionless number dictating the invasion process. It is a generalisation of the fluctuation number introduced by \citet{Maloy2021}, \citet{meheust2002} and \citet{auradou1999}. The quantity $1/N(p_c,x)$ characterises the typical width of the capillary threshold fluctuations, and, in the scenario of interest in our experiments, $a(G-\partial \hat{p}/\partial x)$ characterises the gravitational forces at the pore scale, corrected for the particular structure of the porous matrix. If the material disorder is important relatively to the gradient term, such gradient becomes negligible at small scales (and reciprocally).\\
One can compare $F$ to other dimensionless numbers usually used to predict flow patterns in porous materials. In our case, a relevant one would for instance be the Bond number $Bo$\,\cite{Bond_Number}, which compares the gravitational forces to the capillary ones. Previous works (e.g.,\,\cite{Wilkinson1984,Pride2022}) have proposed imbibition and drainage front widths to indeed scale as $Bo^{-\nu/(1+\nu)}$. Yet, and contrarily to $F$, the Bond number does not offer any insight on the actual material disorder (e.g., $N$) or on an eventual structural trend (e.g., $\partial \hat{p}/\partial x$).\\
Consider now a case without external field ($G=0$). As shown in Fig.\,\ref{gradfig}a, our models were designed so that the unit-less distribution $\tilde{N}(\hat{p}_t/\hat{p}_c(x)) = \hat{p}_c(x) \cdot N(\hat{p}_t, x)$ is conserved along the $x$ direction (i.e., it does not depend on $x$). Additionally, because the pore throats in our model is to scale the same way as the size of the pores $a$, we have, as per the Young-Laplace law, $\hat{p}_c(x)\propto \gamma \cos(\phi)/a(x)$. Finally, we can rewrite Eqs.\,(\ref{eta1}) and\,(\ref{eqF}) as:
\begin{equation}
        \eta\propto\left[\frac{\tilde{N}(1)}{\gamma \cos(\phi)}\left(-\frac{\partial\hat{p}_c(x)}{\partial x}\right)\right]^{-\beta} \times a^{-2\beta+1},
    \label{eqfin}
\end{equation}
where $-2\beta+1$ is close to $0$ so that the effect of the pore size $a$ on the width of the front is small compared to that of the other terms. This last expression is only a particular case of Eqs.\,(\ref{eta1}) and\,(\ref{eqF}), where the front width does not significantly evolve as the invasion progresses. In a more general porous material, the spatial distribution in pore size $a(x)$ and/or in pore invasion threshold $N(p_c, x)$ would matter.

\section{\label{sec:experiments} Experimental results}

\begin{figure}[b]
\centering
\includegraphics[width=1.0\linewidth]{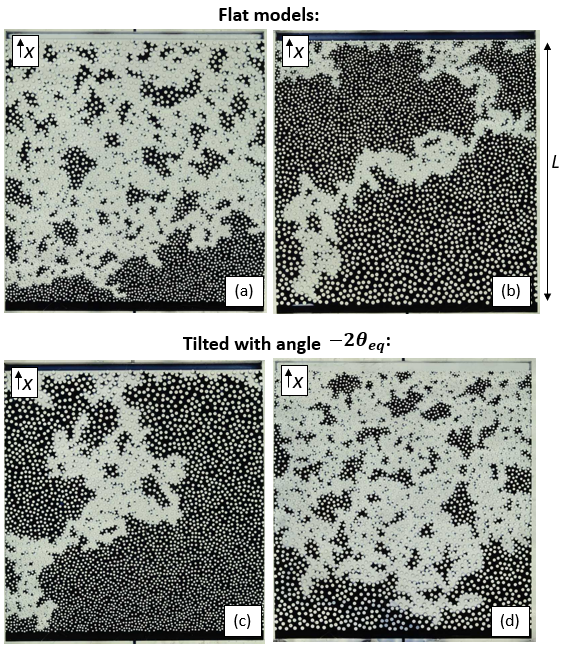}
\caption{Drainage experiments in 3D printed quasi two-dimensional model. Air invaded from the top of the pictures into a glycerine/water solution coloured with nigrosin. In (a), $\partial \hat{p}_c(x)/\partial x <0$, which corresponds to a decrease in `permeability' as the flow progresses (the air invades from the big pores to the small ones). The invasion is stable. In (b), $\partial \hat{p}_c(x)/\partial x > 0$ which corresponds to an increase in `permeability' with the flow progression and an unstable processes. In both case, the model is horizontal ($G=0$). (c) and (d) show the same experiments in chosen destabilising and a stabilising gravity field (respectively) with tilt angle $-2\theta_\text{eq}$ (see Fig.\,\ref{syst} and\,\ref{dist-shift}). The stability of the flow is thus reversed. 
}
\label{gradfig} 
\end{figure}

From Eq.\,(\ref{eqfin}), we see that a negative gradient in the critical capillary pressure threshold will stabilise the front while a positive gradient will make it more unstable. In this sentence, the underlying convention is that the direction defining the gradient is opposing the general flow direction. This is illustrated in Fig. \ref{gradfig} that shows two drainage experiments in such configurations. In Fig.\,\ref{gradfig}a we have a  negative  gradient in the critical capillary pressure threshold and the front is stabilised while in \ref{gradfig}b the gradient is positive (the inlet and outlet having been swapped) and the front is unstable with a  growing  finger. In both case the external (gravity) field is null. The stabilisation/destabilisation of the front is also reflected in the breakthrough time, which was three times smaller in the case with the destabilising gradient. In Fig.\,\ref{gradfig}a the breakthrough time was $60.6$\,hours min while in Fig.\,\ref{gradfig}b it was reduced to $20.2$\,hours (the externally imposed flow rate was the same for both experiments.)\\
In Figs.\,\ref{gradfig}c and\,\ref{gradfig}d, we show the same experiments in a particular gravity field. The model is tilted with an angle $-2\theta_\text{eq}=-2\text{asin}(\frac{-\partial \hat{p}_c(x)}{\partial x}/\Delta\rho g_0)$ as explained in Fig.\,\ref{dist-shift}, which effectively inverses the sign of the $\left(G-\frac{\partial\hat{p}_c(x)}{\partial x}\right)$ term in Eq.\,(\ref{eqF}).\\
We should here restate that Eq.\,(\ref{eqF}) only applies to the cases where the front is stable (i.e.,\, to Figs.\,\ref{gradfig}a and\,\ref{gradfig}d) as front stability was an underlying hypothesis (to write Eq.\,(\ref{peqpc})). As expected, for these two different experiments for which the fluctuation number $F$ is the same, the features of the invasion front are very similar. In the case of unstable fronts, we suggest that, the fluctuation number could characterise the width of the invading fingers, rather than the width of the front. Indeed, in gravitational drainage experiments, such a finger width was indeed proposed (e.g.,\,\cite{vasseur_finger,Frette92}) to scale with the Bond number, with the same exponent $\beta$ which we have here considered. In the case of the drainage of a matrix with a structural gradient (that is here of interest), this hypothesis will be further verified in section\,\ref{sec:desta}, with the help of invasion percolation simulations.\\

\section{\label{sec:simul} Invasion-Percolation simulations}

\subsection{Stabilising gradient}

\begin{figure}
\centering
\includegraphics[width=1.0\linewidth]{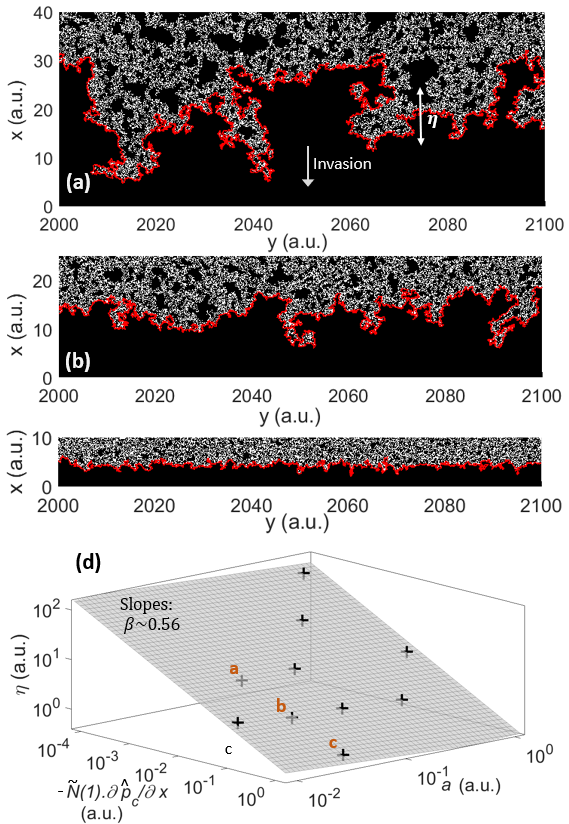}
\caption{(a) to (c): Invasion percolation maps for the simulations denoted in (d), that have different stabilising gradient in invasion threshold. The invading non-wetting phase occupies the whites pixels. The plain (red) line is the drainage front on which the width $\eta$ is computed. (d): Width of the front as a function of the pore size $a$, the gradient $\partial \hat{p_c} / \partial x$ and the local distribution in capillary pressure $\tilde{N}$ (logarithmic scales). Each cross was computed on an independent simulation and the grey plane is a fit of Eq.\,(\ref{eqfin}). The slopes of this plane along the two axes direction are given respectively by $-\beta$ and by $1-2\beta$ and both indicate $\beta\sim0.56$.}
\label{Fig:sims} 
\end{figure}

Fully validating Eq.\,(\ref{eqfin}) with experimental results remained a challenge. Indeed, because of the maximum size ($\sim 30$\,cm) of the models we could print, and because of the minimum distance between printed grains (a fraction of millimetre) before they tended to unexpectedly merge, the range of gradient in capillary pressure that could be investigated was small. Due to the theoretical scaling between such gradient and the front width (i.e., Eq.\,(\ref{eqfin})), the range in obtainable $\eta$ was even smaller. Therefore, to verify our theoretical framework, we ran some invasion-percolation simulations, known to represent well capillary invasion processes.\\
These simulations are performed on square lattices (indexed along a $x$ and $y$ directions), which abide to the same conditions that underlie Eq.\,(\ref{eqfin}). The distribution $N$ in invasion threshold $\hat{p}_t$ for the pixels at a given $x$ is random and uniformly distributed. Similarly to Fig.\,\ref{dist-shift} the mean values of these distributions follow an arbitrary gradient along $x$ but the distributions' width is such that $\tilde{N}(p_t/\hat{p_c})$ is conserved along $x$. Additionally, the size $a(x)$ of a pixel scales as $1/\hat{p_c}(x)$. Here, because our matrices' connectivity is 4, the percolation pressure $\hat{p}_c$ is such that $\int_{0}^{\hat{p}_c} N(\hat{p}_t) \,\mathrm{d}\hat{p}_t = 0.5$ (e.g., see\,\cite{moura2015,perco_theo}).\\
At the initial stage, solely the first line of the matrix (the inlet) is invaded. At each time step, one new pixel is invaded. This pixel is the neighbour of the invading phase which has the lowest invasion threshold and is still connected to the matrix outlet. We ran ten simulations in the stable invasion domain ($\partial \hat{p_c} / \partial x <0$), varying the value of the gradient, but also the average size of the pixel $a$ in the model and the local width of the distribution in invasion threshold $1/\tilde{N(1)}$. We thus covered four decades of $\tilde{N}(1)\times\partial \hat{p_c} / \partial x <0$ and two decades of $a$. For each simulation, we extracted the simulated front width $\eta$, once it reached a plateau. We defined this width as the standard deviation of the $x$ coordinate of the front between the invading phase and the main cluster of the defending one (see Fig.\,\ref{Fig:sims}).\\
Finally, with a least squares method, we fitted Eq.\,(\ref{eqfin}) to the obtained data in order to invert for the $\beta$ exponent. This procedure provided a good fit of the simulated data with a coefficient of determination $R^2\sim0.998$, and we found $\beta\sim0.56\pm0.02$, where the accuracy of the fit is computed by letting $R^2$ vary by $5$\%. Such value for $\beta$ is close to our theoretical prediction for percolation theory.

\subsection{\label{sec:desta} Destabilising gradient}

\begin{figure}[b]
\centering
\includegraphics[width=1.0\linewidth]{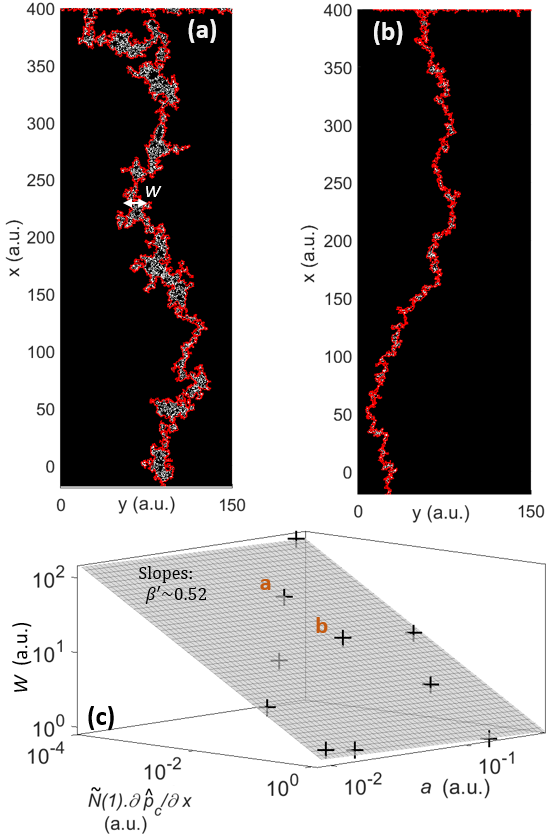}
\caption{(a) and (b): Invasion percolation maps for the simulations denoted in (c), that have different destabilising gradient in invasion threshold. The invading non-wetting phase occupies the whites pixels. The plain (red) line is the drainage front on which the width $W$ is computed. (d): Width of the finger as a function of the pore size $a$, the gradient $\partial \hat{p_c} / \partial x$ and the local distribution in capillary pressure $\tilde{N}$ (logarithmic scales). Each cross was computed on an independent simulation and the grey plane is a fit of Eq.\,(\ref{eqfin2}). The slopes of this plane along the two axes direction are given respectively by $-\beta'$ and by $1-2\beta'$ and both indicate $\beta'\sim0.52$.}
\label{Fig:simsuns} 
\end{figure}

We also ran similar simulations, but in the unstable case, that is with $\partial \hat{p_c} / \partial x >0$. We varied the same parameters (i.e., $\partial \hat{p_c} / \partial x$, $\tilde{N}$ and $a$). In this unstable configuration, rather than characterising the width of a stable front, we characterised the average width $W$ of the growing invasion finger. We defined such width as $W=A/l$, where $A$ is the area occupied by the finger (i.e., the area surrounded by the red lines in Fig.\,\ref{Fig:simsuns}, where the results are shown), and where $l$ is the length of the finger along the $x$ direction. Comparing Figs.\,\ref{Fig:sims} and\,\ref{Fig:simsuns}, one can notice how similar is the scaling of $W$ and $\eta$. Although Eq.\,(\ref{eqfin}) was only formally derived for stable fronts, we then analogously write for the width of the fingers:
\begin{equation}
        W\propto\left[\frac{\tilde{N}(1)}{\gamma \cos(\phi)}\left(+\frac{\partial\hat{p}_c(x)}{\partial x}\right)\right]^{-\beta'} \times a^{-2\beta'+1},
    \label{eqfin2}
\end{equation}
when the same assumptions for the matrix structure than those underlying Eq.\,(\ref{eqfin}) are respected.
Fitting this expression to our simulations' results, we obtained a good match ($R^2\sim0.992$) for $\beta'\sim 0.52\pm0.05$. This value is close to the value of $\beta$, and a similar scaling of $W$ with respect to the Bond number (rather than to $F$) was reported in experimental observations of unstable drainage fingers growing in a gravitational field (e.g.\,\cite{vasseur_finger,Frette92}).

\section{\label{sec:conclusion} Conclusion}

In this paper, we discussed the importance of capillary fluctuations in porous media, as well as the characteristic length scales in two-phase flow patterns set by the competition between capillary fluctuations, external fields (e.g., gravitational or viscous ones) and a gradient in the matrix percolation pressure.\\
In the case of fluid fronts that are stabilised by these fields and porous media geometry, the fluctuation number $F$, which describes the scaling of the front width $\eta$, was introduced. There, the derived scaling exponents directly result from percolation theory. When considering a viscous and gravitational field, the theory describes well the scaling of the width of the fluid front and the final saturation of the fluid left behind the invasion front observed in laboratory experiments\,\cite{Maloy2021}. As shown here, it can also predict the stabilisation and destabilisation of the front width in such experiments depending on the sign of the spatial gradient in the critical capillary pressure. Truly, more experiments are needed to conduct a quantitative experimental investigation of the dependence of the scaling of the front width $\eta$ when a structural gradient is present, the challenge being to obtain a permeability gradient varying over several decades in the laboratory. Standard invasion-percolation simulations have however here underlined how reasonable is the predicted scaling law. In the case of a destabilised flow, these simulations also allowed to infer a similar scaling law for the width of growing drainage fingers.\\
On a length scale smaller than $\eta$, the structure within the front is generally fractal, while on a length scale larger than $\eta$, it is homogeneous. The characteristic length scale $\eta$ should thus be of primary importance in defining a relevant Representative Elementary Volume (REV) for an average Darcy description of the two-phase flow problem \cite{ayaz2020}. We suggest that characterising the fluctuation number $F$, for instance from drilled core samples in geological contexts, would help in this prediction of the front width. One could also extend the approach of \citet{moura2015} and \citet{ayaz2020} for systems with structural gradients, thus characterising the relationship between the fluids' pressures and saturations. A good knowledge of this relationship is indeed relevant for reservoir geophysicists and hydrologists.\\
In these geological applications, one should also consider the usual repeating sequences of sediment layers with a gradient in permeability in a given direction. This is for instance often observed in fluvial or turbidite deposition\,\cite{Graded_bedding}. There, if the gradient inside a single unit is stabilising, a change in layer would yet correspond to a local but brutal destabilising effect (and reciprocally). Predicting the flow behaviour along large distances would thus likely require to take into consideration the typical wave lengths of such layer repetitions, and compare them to the typical length scale $\eta$ of the invading pattern. Such a study would be a natural continuation to the present work. 

\section*{Authors contributions and acknowledgements}

\noindent KJM and TVD proposed the theory developed in this article, TVD and MM performed the experimental work, TVD wrote and ran the numerical simulations, and RT advised in both the theory and the numerical implementation. TVD and KJM wrote the first version of the manuscript and all the authors agreed on the submitted version.\\
We acknowledge the support of the University of Oslo, of the Njord Center, and of SFF Porelab (project number 262644 of the Research Council of Norway). We also thank the IRP France-Norway D-FFRACT.\\
We declare no competing interest in the publishing of this work. A funding support from the University of Strasbourg is acknowledged. Readers are welcome to comment and correspondence should be addressed to maloy@fys.uio.no.

\bibliography{sources}

\begin{thebibliography}{30}%
\makeatletter
\providecommand \@ifxundefined [1]{%
 \@ifx{#1\undefined}
}%
\providecommand \@ifnum [1]{%
 \ifnum #1\expandafter \@firstoftwo
 \else \expandafter \@secondoftwo
 \fi
}%
\providecommand \@ifx [1]{%
 \ifx #1\expandafter \@firstoftwo
 \else \expandafter \@secondoftwo
 \fi
}%
\providecommand \natexlab [1]{#1}%
\providecommand \enquote  [1]{``#1''}%
\providecommand \bibnamefont  [1]{#1}%
\providecommand \bibfnamefont [1]{#1}%
\providecommand \citenamefont [1]{#1}%
\providecommand \href@noop [0]{\@secondoftwo}%
\providecommand \href [0]{\begingroup \@sanitize@url \@href}%
\providecommand \@href[1]{\@@startlink{#1}\@@href}%
\providecommand \@@href[1]{\endgroup#1\@@endlink}%
\providecommand \@sanitize@url [0]{\catcode `\\12\catcode `\$12\catcode
  `\&12\catcode `\#12\catcode `\^12\catcode `\_12\catcode `\%12\relax}%
\providecommand \@@startlink[1]{}%
\providecommand \@@endlink[0]{}%
\providecommand \url  [0]{\begingroup\@sanitize@url \@url }%
\providecommand \@url [1]{\endgroup\@href {#1}{\urlprefix }}%
\providecommand \urlprefix  [0]{URL }%
\providecommand \Eprint [0]{\href }%
\providecommand \doibase [0]{http://dx.doi.org/}%
\providecommand \selectlanguage [0]{\@gobble}%
\providecommand \bibinfo  [0]{\@secondoftwo}%
\providecommand \bibfield  [0]{\@secondoftwo}%
\providecommand \translation [1]{[#1]}%
\providecommand \BibitemOpen [0]{}%
\providecommand \bibitemStop [0]{}%
\providecommand \bibitemNoStop [0]{.\EOS\space}%
\providecommand \EOS [0]{\spacefactor3000\relax}%
\providecommand \BibitemShut  [1]{\csname bibitem#1\endcsname}%
\let\auto@bib@innerbib\@empty
\bibitem [{\citenamefont {Lenormand}\ \emph {et~al.}(1988)\citenamefont
  {Lenormand}, \citenamefont {Touboul},\ and\ \citenamefont
  {Zaarcone}}]{lenormand1988}%
  \BibitemOpen
  \bibfield  {author} {\bibinfo {author} {\bibfnamefont {R.}~\bibnamefont
  {Lenormand}}, \bibinfo {author} {\bibfnamefont {E.}~\bibnamefont {Touboul}},
  \ and\ \bibinfo {author} {\bibfnamefont {C.}~\bibnamefont {Zaarcone}},\
  }\href@noop {} {\bibfield  {journal} {\bibinfo  {journal} {Journal of Fluid
  Mechanics}\ }\textbf {\bibinfo {volume} {189}},\ \bibinfo {pages} {165}
  (\bibinfo {year} {1988})}\BibitemShut {NoStop}%
\bibitem [{\citenamefont {Lenormand}(1989)}]{lenormand1989}%
  \BibitemOpen
  \bibfield  {author} {\bibinfo {author} {\bibfnamefont {R.}~\bibnamefont
  {Lenormand}},\ }\href@noop {} {\bibfield  {journal} {\bibinfo  {journal}
  {Proceedings of the Royal Society of London A}\ }\textbf {\bibinfo {volume}
  {423}},\ \bibinfo {pages} {159} (\bibinfo {year} {1989})}\BibitemShut
  {NoStop}%
\bibitem [{\citenamefont {Zhao}\ \emph {et~al.}(2016)\citenamefont {Zhao},
  \citenamefont {Mac~Minn},\ and\ \citenamefont {Juanes}}]{zhao2016}%
  \BibitemOpen
  \bibfield  {author} {\bibinfo {author} {\bibfnamefont {B.}~\bibnamefont
  {Zhao}}, \bibinfo {author} {\bibfnamefont {C.~W.}\ \bibnamefont {Mac~Minn}},
  \ and\ \bibinfo {author} {\bibfnamefont {R.}~\bibnamefont {Juanes}},\ }\href
  {\doibase 10.1073/pnas.1603387113} {\bibfield  {journal} {\bibinfo  {journal}
  {PNAS}\ }\textbf {\bibinfo {volume} {113}},\ \bibinfo {pages} {10251}
  (\bibinfo {year} {2016})}\BibitemShut {NoStop}%
\bibitem [{\citenamefont {M\aa{}l\o{}y}\ \emph {et~al.}(1985)\citenamefont
  {M\aa{}l\o{}y}, \citenamefont {Feder},\ and\ \citenamefont
  {J\o{}ssang}}]{maloy1985}%
  \BibitemOpen
  \bibfield  {author} {\bibinfo {author} {\bibfnamefont {K.~J.}\ \bibnamefont
  {M\aa{}l\o{}y}}, \bibinfo {author} {\bibfnamefont {J.}~\bibnamefont {Feder}},
  \ and\ \bibinfo {author} {\bibfnamefont {T.}~\bibnamefont {J\o{}ssang}},\
  }\href@noop {} {\bibfield  {journal} {\bibinfo  {journal} {Physical Review
  Letters}\ }\textbf {\bibinfo {volume} {55}},\ \bibinfo {pages} {2688–}
  (\bibinfo {year} {1985})}\BibitemShut {NoStop}%
\bibitem [{\citenamefont {Chen}\ and\ \citenamefont
  {Wilkinson}(1985)}]{chen1985}%
  \BibitemOpen
  \bibfield  {author} {\bibinfo {author} {\bibfnamefont {J.-D.}\ \bibnamefont
  {Chen}}\ and\ \bibinfo {author} {\bibfnamefont {D.}~\bibnamefont
  {Wilkinson}},\ }\href {\doibase 10.1103/PhysRevLett.55.1892} {\bibfield
  {journal} {\bibinfo  {journal} {Physical Review Letters}\ }\textbf {\bibinfo
  {volume} {55}},\ \bibinfo {pages} {1892} (\bibinfo {year}
  {1985})}\BibitemShut {NoStop}%
\bibitem [{\citenamefont {Mandelbrot}(1982)}]{mandelbrot1982}%
  \BibitemOpen
  \bibfield  {author} {\bibinfo {author} {\bibfnamefont {B.~B.}\ \bibnamefont
  {Mandelbrot}},\ }\href@noop {} {\bibfield  {journal} {\bibinfo  {journal} {W.
  H. Freeman. San Fransisco}\ } (\bibinfo {year} {1982})}\BibitemShut {NoStop}%
\bibitem [{\citenamefont {Feder}(1988)}]{feder1988}%
  \BibitemOpen
  \bibfield  {author} {\bibinfo {author} {\bibfnamefont {J.}~\bibnamefont
  {Feder}},\ }\href@noop {} {\bibfield  {journal} {\bibinfo  {journal} {Plenum,
  New York}\ } (\bibinfo {year} {1988})}\BibitemShut {NoStop}%
\bibitem [{\citenamefont {Weitz}\ \emph {et~al.}(1987)\citenamefont {Weitz},
  \citenamefont {Stokes}, \citenamefont {Ball},\ and\ \citenamefont
  {Kushnick}}]{weitz1987}%
  \BibitemOpen
  \bibfield  {author} {\bibinfo {author} {\bibfnamefont {D.}~\bibnamefont
  {Weitz}}, \bibinfo {author} {\bibfnamefont {J.}~\bibnamefont {Stokes}},
  \bibinfo {author} {\bibfnamefont {R.}~\bibnamefont {Ball}}, \ and\ \bibinfo
  {author} {\bibfnamefont {A.}~\bibnamefont {Kushnick}},\ }\href {\doibase
  10.1103/PhysRevLett.59.2967} {\bibfield  {journal} {\bibinfo  {journal}
  {Physical Review Letters}\ }\textbf {\bibinfo {volume} {59}},\ \bibinfo
  {pages} {2967} (\bibinfo {year} {1987})}\BibitemShut {NoStop}%
\bibitem [{\citenamefont {L{\o}voll}\ \emph {et~al.}(2004)\citenamefont
  {L{\o}voll}, \citenamefont {M{\'e}heust}, \citenamefont {Toussaint},
  \citenamefont {Schmittbuhl},\ and\ \citenamefont
  {M{\aa}l{\o}y}}]{lovoll2004}%
  \BibitemOpen
  \bibfield  {author} {\bibinfo {author} {\bibfnamefont {G.}~\bibnamefont
  {L{\o}voll}}, \bibinfo {author} {\bibfnamefont {Y.}~\bibnamefont
  {M{\'e}heust}}, \bibinfo {author} {\bibfnamefont {R.}~\bibnamefont
  {Toussaint}}, \bibinfo {author} {\bibfnamefont {J.}~\bibnamefont
  {Schmittbuhl}}, \ and\ \bibinfo {author} {\bibfnamefont {K.~J.}\ \bibnamefont
  {M{\aa}l{\o}y}},\ }\href {\doibase 10.1103/PhysRevE.70.026301} {\bibfield
  {journal} {\bibinfo  {journal} {Physical Review E}\ }\textbf {\bibinfo
  {volume} {70}},\ \bibinfo {pages} {026301} (\bibinfo {year}
  {2004})}\BibitemShut {NoStop}%
\bibitem [{\citenamefont {Måløy}\ \emph {et~al.}(2021)\citenamefont
  {Måløy}, \citenamefont {Moura}, \citenamefont {Hansen}, \citenamefont
  {Flekkøy},\ and\ \citenamefont {Toussaint}}]{Maloy2021}%
  \BibitemOpen
  \bibfield  {author} {\bibinfo {author} {\bibfnamefont {K.~J.}\ \bibnamefont
  {Måløy}}, \bibinfo {author} {\bibfnamefont {M.}~\bibnamefont {Moura}},
  \bibinfo {author} {\bibfnamefont {A.}~\bibnamefont {Hansen}}, \bibinfo
  {author} {\bibfnamefont {E.}~\bibnamefont {Flekkøy}}, \ and\ \bibinfo
  {author} {\bibfnamefont {R.}~\bibnamefont {Toussaint}},\ }\href {\doibase
  10.3389/fphy.2021.796019} {\bibfield  {journal} {\bibinfo  {journal}
  {Frontiers in Physics}\ }\textbf {\bibinfo {volume} {9}} (\bibinfo {year}
  {2021}),\ 10.3389/fphy.2021.796019}\BibitemShut {NoStop}%
\bibitem [{\citenamefont {Lenormand}\ and\ \citenamefont
  {Zarcone}(1985)}]{lenormand1985}%
  \BibitemOpen
  \bibfield  {author} {\bibinfo {author} {\bibfnamefont {R.}~\bibnamefont
  {Lenormand}}\ and\ \bibinfo {author} {\bibfnamefont {C.}~\bibnamefont
  {Zarcone}},\ }\href {\doibase 10.1103/PhysRevLett.54.2226} {\bibfield
  {journal} {\bibinfo  {journal} {Physical Review Letters}\ }\textbf {\bibinfo
  {volume} {54}},\ \bibinfo {pages} {2226} (\bibinfo {year}
  {1985})}\BibitemShut {NoStop}%
\bibitem [{\citenamefont {Primkulov}\ \emph {et~al.}(2022)\citenamefont
  {Primkulov}, \citenamefont {Zhao}, \citenamefont {MacMinn},\ and\
  \citenamefont {Juanes}}]{aval_imb}%
  \BibitemOpen
  \bibfield  {author} {\bibinfo {author} {\bibfnamefont {B.~K.}\ \bibnamefont
  {Primkulov}}, \bibinfo {author} {\bibfnamefont {B.}~\bibnamefont {Zhao}},
  \bibinfo {author} {\bibfnamefont {C.~W.}\ \bibnamefont {MacMinn}}, \ and\
  \bibinfo {author} {\bibfnamefont {R.}~\bibnamefont {Juanes}},\ }\href
  {\doibase 10.1038/s42005-022-00826-1} {\bibfield  {journal} {\bibinfo
  {journal} {Communications Physics}\ }\textbf {\bibinfo {volume} {5}},\
  \bibinfo {pages} {52} (\bibinfo {year} {2022})}\BibitemShut {NoStop}%
\bibitem [{\citenamefont {Wilkinson}(1984)}]{Wilkinson1984}%
  \BibitemOpen
  \bibfield  {author} {\bibinfo {author} {\bibfnamefont {D.}~\bibnamefont
  {Wilkinson}},\ }\href {\doibase 10.1103/PhysRevA.30.520} {\bibfield
  {journal} {\bibinfo  {journal} {Phys. Rev. A}\ }\textbf {\bibinfo {volume}
  {30}},\ \bibinfo {pages} {520} (\bibinfo {year} {1984})}\BibitemShut
  {NoStop}%
\bibitem [{\citenamefont {Birovljev}\ \emph {et~al.}(1991)\citenamefont
  {Birovljev}, \citenamefont {Furuberg}, \citenamefont {Feder}, \citenamefont
  {J{\o}ssang}, \citenamefont {M{\aa}l{\o}y},\ and\ \citenamefont
  {Aharony}}]{birovljev91}%
  \BibitemOpen
  \bibfield  {author} {\bibinfo {author} {\bibfnamefont {A.}~\bibnamefont
  {Birovljev}}, \bibinfo {author} {\bibfnamefont {L.}~\bibnamefont {Furuberg}},
  \bibinfo {author} {\bibfnamefont {J.}~\bibnamefont {Feder}}, \bibinfo
  {author} {\bibfnamefont {T.}~\bibnamefont {J{\o}ssang}}, \bibinfo {author}
  {\bibfnamefont {K.~J.}\ \bibnamefont {M{\aa}l{\o}y}}, \ and\ \bibinfo
  {author} {\bibfnamefont {A.}~\bibnamefont {Aharony}},\ }\href@noop {}
  {\bibfield  {journal} {\bibinfo  {journal} {Physical Review Letters}\
  }\textbf {\bibinfo {volume} {67}},\ \bibinfo {pages} {584} (\bibinfo {year}
  {1991})}\BibitemShut {NoStop}%
\bibitem [{\citenamefont {Frette}\ \emph {et~al.}(1992)\citenamefont {Frette},
  \citenamefont {Feder}, \citenamefont {J\o{}ssang},\ and\ \citenamefont
  {Meakin}}]{Frette92}%
  \BibitemOpen
  \bibfield  {author} {\bibinfo {author} {\bibfnamefont {V.}~\bibnamefont
  {Frette}}, \bibinfo {author} {\bibfnamefont {J.}~\bibnamefont {Feder}},
  \bibinfo {author} {\bibfnamefont {T.}~\bibnamefont {J\o{}ssang}}, \ and\
  \bibinfo {author} {\bibfnamefont {P.}~\bibnamefont {Meakin}},\ }\href
  {\doibase 10.1103/PhysRevLett.68.3164} {\bibfield  {journal} {\bibinfo
  {journal} {Physical Review Letters}\ }\textbf {\bibinfo {volume} {68}},\
  \bibinfo {pages} {3164} (\bibinfo {year} {1992})}\BibitemShut {NoStop}%
\bibitem [{\citenamefont {Auradou}\ \emph {et~al.}(1999)\citenamefont
  {Auradou}, \citenamefont {M\aa{}l\o{}y}, \citenamefont {Schmittbuhl},
  \citenamefont {Hansen},\ and\ \citenamefont {Bideau}}]{auradou1999}%
  \BibitemOpen
  \bibfield  {author} {\bibinfo {author} {\bibfnamefont {H.}~\bibnamefont
  {Auradou}}, \bibinfo {author} {\bibfnamefont {K.~J.}\ \bibnamefont
  {M\aa{}l\o{}y}}, \bibinfo {author} {\bibfnamefont {J.}~\bibnamefont
  {Schmittbuhl}}, \bibinfo {author} {\bibfnamefont {A.}~\bibnamefont {Hansen}},
  \ and\ \bibinfo {author} {\bibfnamefont {D.}~\bibnamefont {Bideau}},\ }\href
  {\doibase 10.1103/PhysRevE.60.7224} {\bibfield  {journal} {\bibinfo
  {journal} {Physical Review E}\ }\textbf {\bibinfo {volume} {60}},\ \bibinfo
  {pages} {7224} (\bibinfo {year} {1999})}\BibitemShut {NoStop}%
\bibitem [{\citenamefont {Méheust}\ \emph {et~al.}(2002)\citenamefont
  {Méheust}, \citenamefont {L{\o}voll}, \citenamefont {M{\aa}l{\o}y},\ and\
  \citenamefont {Schmittbuhl}}]{meheust2002}%
  \BibitemOpen
  \bibfield  {author} {\bibinfo {author} {\bibfnamefont {Y.}~\bibnamefont
  {Méheust}}, \bibinfo {author} {\bibfnamefont {G.}~\bibnamefont {L{\o}voll}},
  \bibinfo {author} {\bibfnamefont {K.~J.}\ \bibnamefont {M{\aa}l{\o}y}}, \
  and\ \bibinfo {author} {\bibfnamefont {J.}~\bibnamefont {Schmittbuhl}},\
  }\href@noop {} {\bibfield  {journal} {\bibinfo  {journal} {Physical Review
  E}\ }\textbf {\bibinfo {volume} {66}},\ \bibinfo {pages} {051603} (\bibinfo
  {year} {2002})}\BibitemShut {NoStop}%
\bibitem [{\citenamefont {Breen}\ \emph {et~al.}(2022)\citenamefont {Breen},
  \citenamefont {Pride}, \citenamefont {Masson},\ and\ \citenamefont
  {Manga}}]{Pride2022}%
  \BibitemOpen
  \bibfield  {author} {\bibinfo {author} {\bibfnamefont {S.~J.}\ \bibnamefont
  {Breen}}, \bibinfo {author} {\bibfnamefont {S.~R.}\ \bibnamefont {Pride}},
  \bibinfo {author} {\bibfnamefont {Y.}~\bibnamefont {Masson}}, \ and\ \bibinfo
  {author} {\bibfnamefont {M.}~\bibnamefont {Manga}},\ }\href {\doibase
  https://doi.org/10.1016/j.advwatres.2022.104150} {\bibfield  {journal}
  {\bibinfo  {journal} {Advances in Water Resources}\ }\textbf {\bibinfo
  {volume} {162}},\ \bibinfo {pages} {104150} (\bibinfo {year}
  {2022})}\BibitemShut {NoStop}%
\bibitem [{\citenamefont {Toussaint}\ \emph {et~al.}(2012)\citenamefont
  {Toussaint}, \citenamefont {M{\aa}l{\o}y}, \citenamefont {M{\'e}heust},
  \citenamefont {L{\o}voll}, \citenamefont {Jankov}, \citenamefont
  {Sch{\"a}fer},\ and\ \citenamefont {Schmittbuhl}}]{toussaint2012}%
  \BibitemOpen
  \bibfield  {author} {\bibinfo {author} {\bibfnamefont {R.}~\bibnamefont
  {Toussaint}}, \bibinfo {author} {\bibfnamefont {K.~J.}\ \bibnamefont
  {M{\aa}l{\o}y}}, \bibinfo {author} {\bibfnamefont {Y.}~\bibnamefont
  {M{\'e}heust}}, \bibinfo {author} {\bibfnamefont {G.}~\bibnamefont
  {L{\o}voll}}, \bibinfo {author} {\bibfnamefont {M.}~\bibnamefont {Jankov}},
  \bibinfo {author} {\bibfnamefont {G.}~\bibnamefont {Sch{\"a}fer}}, \ and\
  \bibinfo {author} {\bibfnamefont {J.}~\bibnamefont {Schmittbuhl}},\
  }\href@noop {} {\bibfield  {journal} {\bibinfo  {journal} {Vadose Zone
  Journal}\ }\textbf {\bibinfo {volume} {11}} (\bibinfo {year}
  {2012})}\BibitemShut {NoStop}%
\bibitem [{\citenamefont {Rabbani}\ \emph {et~al.}(2018)\citenamefont
  {Rabbani}, \citenamefont {Or}, \citenamefont {Liu}, \citenamefont {Lai},
  \citenamefont {Lu}, \citenamefont {Datta},\ and\ \citenamefont
  {Stone}}]{rabbani2018}%
  \BibitemOpen
  \bibfield  {author} {\bibinfo {author} {\bibfnamefont {H.~S.}\ \bibnamefont
  {Rabbani}}, \bibinfo {author} {\bibfnamefont {D.}~\bibnamefont {Or}},
  \bibinfo {author} {\bibfnamefont {Y.}~\bibnamefont {Liu}}, \bibinfo {author}
  {\bibfnamefont {C.-Y.}\ \bibnamefont {Lai}}, \bibinfo {author} {\bibfnamefont
  {N.~B.}\ \bibnamefont {Lu}}, \bibinfo {author} {\bibfnamefont {S.~S.}\
  \bibnamefont {Datta}}, \ and\ \bibinfo {author} {\bibfnamefont {H.~A.}\
  \bibnamefont {Stone}},\ }\href@noop {} {\bibfield  {journal} {\bibinfo
  {journal} {PNAS}\ }\textbf {\bibinfo {volume} {115}},\ \bibinfo {pages}
  {4833} (\bibinfo {year} {2018})}\BibitemShut {NoStop}%
\bibitem [{\citenamefont {Lu}\ \emph {et~al.}(2019)\citenamefont {Lu},
  \citenamefont {Browne},\ and\ \citenamefont {Amchin}}]{lu2019}%
  \BibitemOpen
  \bibfield  {author} {\bibinfo {author} {\bibfnamefont {N.~B.}\ \bibnamefont
  {Lu}}, \bibinfo {author} {\bibfnamefont {C.~A.}\ \bibnamefont {Browne}}, \
  and\ \bibinfo {author} {\bibfnamefont {D.~B.}\ \bibnamefont {Amchin}},\
  }\href {\doibase 10.1103/PhysRevFluids.4.084303} {\bibfield  {journal}
  {\bibinfo  {journal} {Physical Review Fluids}\ }\textbf {\bibinfo {volume}
  {4}},\ \bibinfo {pages} {084303} (\bibinfo {year} {2019})}\BibitemShut
  {NoStop}%
\bibitem [{\citenamefont {Kuenen}(1953)}]{Graded_bedding}%
  \BibitemOpen
  \bibfield  {author} {\bibinfo {author} {\bibfnamefont {P.~H.}\ \bibnamefont
  {Kuenen}},\ }\href {\doibase 10.1306/5CEADCC8-16BB-11D7-8645000102C1865D}
  {\bibfield  {journal} {\bibinfo  {journal} {AAPG Bulletin}\ }\textbf
  {\bibinfo {volume} {37}},\ \bibinfo {pages} {1044} (\bibinfo {year}
  {1953})}\BibitemShut {NoStop}%
\bibitem [{\citenamefont {Moura}\ \emph {et~al.}(2015)\citenamefont {Moura},
  \citenamefont {Florentino}, \citenamefont {M{\aa}l{\o}y}, \citenamefont
  {Schafer},\ and\ \citenamefont {Toussaint}}]{moura2015}%
  \BibitemOpen
  \bibfield  {author} {\bibinfo {author} {\bibfnamefont {M.}~\bibnamefont
  {Moura}}, \bibinfo {author} {\bibfnamefont {E.}~\bibnamefont {Florentino}},
  \bibinfo {author} {\bibfnamefont {K.}~\bibnamefont {M{\aa}l{\o}y}}, \bibinfo
  {author} {\bibfnamefont {G.}~\bibnamefont {Schafer}}, \ and\ \bibinfo
  {author} {\bibfnamefont {R.}~\bibnamefont {Toussaint}},\ }\href {\doibase
  doi:10.1002/2015WR017196} {\bibfield  {journal} {\bibinfo  {journal} {Water
  Resources Research}\ }\textbf {\bibinfo {volume} {51}} (\bibinfo {year}
  {2015}),\ doi:10.1002/2015WR017196}\BibitemShut {NoStop}%
\bibitem [{\citenamefont {Sauffer}(1994)}]{perco_theo}%
  \BibitemOpen
  \bibfield  {author} {\bibinfo {author} {\bibfnamefont {D.}~\bibnamefont
  {Sauffer}},\ }\href@noop {} {\emph {\bibinfo {title} {Introduction to
  percolation theory}}}\ (\bibinfo  {publisher} {Taylor \& Francis},\ \bibinfo
  {address} {London},\ \bibinfo {year} {1994})\BibitemShut {NoStop}%
\bibitem [{For(2020)}]{Formlabs}%
  \BibitemOpen
  \href
  {https://media.formlabs.com/m/5f0ed4f707037528/original/-ENUS-Form-3L-Manual.pdf}
  {\emph {\bibinfo {title} {Technical Information, {Form 3L}}}},\ \bibinfo
  {type} {Tech. Rep.}\ (\bibinfo  {institution} {Formlabs},\ \bibinfo {year}
  {2020})\BibitemShut {NoStop}%
\bibitem [{\citenamefont {Hinrichsen}\ \emph {et~al.}(1986)\citenamefont
  {Hinrichsen}, \citenamefont {Feder},\ and\ \citenamefont
  {J{\o}ssang}}]{hinrichsen1986}%
  \BibitemOpen
  \bibfield  {author} {\bibinfo {author} {\bibfnamefont {E.~L.}\ \bibnamefont
  {Hinrichsen}}, \bibinfo {author} {\bibfnamefont {J.}~\bibnamefont {Feder}}, \
  and\ \bibinfo {author} {\bibfnamefont {T.}~\bibnamefont {J{\o}ssang}},\
  }\href@noop {} {\bibfield  {journal} {\bibinfo  {journal} {Jornal of
  Statistical Physics}\ }\textbf {\bibinfo {volume} {44}},\ \bibinfo {pages}
  {793} (\bibinfo {year} {1986})}\BibitemShut {NoStop}%
\bibitem [{\citenamefont {Sapoval}\ \emph {et~al.}(1985)\citenamefont
  {Sapoval}, \citenamefont {Rosso},\ and\ \citenamefont {Gouyet}}]{sapova1985}%
  \BibitemOpen
  \bibfield  {author} {\bibinfo {author} {\bibfnamefont {E.}~\bibnamefont
  {Sapoval}}, \bibinfo {author} {\bibfnamefont {M.}~\bibnamefont {Rosso}}, \
  and\ \bibinfo {author} {\bibfnamefont {J.}~\bibnamefont {Gouyet}},\
  }\href@noop {} {\bibfield  {journal} {\bibinfo  {journal} {Journal de
  Physique Lettres}\ }\textbf {\bibinfo {volume} {46}},\ \bibinfo {pages}
  {L149} (\bibinfo {year} {1985})}\BibitemShut {NoStop}%
\bibitem [{\citenamefont {Hager}(2012)}]{Bond_Number}%
  \BibitemOpen
  \bibfield  {author} {\bibinfo {author} {\bibfnamefont {W.~H.}\ \bibnamefont
  {Hager}},\ }\href {\doibase 10.1080/00221686.2011.649839} {\bibfield
  {journal} {\bibinfo  {journal} {Journal of Hydraulic Research}\ }\textbf
  {\bibinfo {volume} {50}},\ \bibinfo {pages} {3} (\bibinfo {year}
  {2012})}\BibitemShut {NoStop}%
\bibitem [{\citenamefont {Vasseur}\ \emph {et~al.}(2013)\citenamefont
  {Vasseur}, \citenamefont {Luo}, \citenamefont {Yan}, \citenamefont {Loggia},
  \citenamefont {Toussaint},\ and\ \citenamefont
  {Schmittbuhl}}]{vasseur_finger}%
  \BibitemOpen
  \bibfield  {author} {\bibinfo {author} {\bibfnamefont {G.}~\bibnamefont
  {Vasseur}}, \bibinfo {author} {\bibfnamefont {X.}~\bibnamefont {Luo}},
  \bibinfo {author} {\bibfnamefont {J.}~\bibnamefont {Yan}}, \bibinfo {author}
  {\bibfnamefont {D.}~\bibnamefont {Loggia}}, \bibinfo {author} {\bibfnamefont
  {R.}~\bibnamefont {Toussaint}}, \ and\ \bibinfo {author} {\bibfnamefont
  {J.}~\bibnamefont {Schmittbuhl}},\ }\href {\doibase
  https://doi.org/10.1016/j.marpetgeo.2013.04.020} {\bibfield  {journal}
  {\bibinfo  {journal} {Marine and Petroleum Geology}\ }\textbf {\bibinfo
  {volume} {45}},\ \bibinfo {pages} {150} (\bibinfo {year} {2013})}\BibitemShut
  {NoStop}%
\bibitem [{\citenamefont {Ayaz}\ \emph {et~al.}(2020)\citenamefont {Ayaz},
  \citenamefont {Toussaint}, \citenamefont {Schafer},\ and\ \citenamefont
  {M{\aa}l{\o}y}}]{ayaz2020}%
  \BibitemOpen
  \bibfield  {author} {\bibinfo {author} {\bibfnamefont {M.}~\bibnamefont
  {Ayaz}}, \bibinfo {author} {\bibfnamefont {R.}~\bibnamefont {Toussaint}},
  \bibinfo {author} {\bibfnamefont {G.}~\bibnamefont {Schafer}}, \ and\
  \bibinfo {author} {\bibfnamefont {K.~J.}\ \bibnamefont {M{\aa}l{\o}y}},\
  }\href {\doibase e2019WR026279} {\bibfield  {journal} {\bibinfo  {journal}
  {Water Resources Research}\ }\textbf {\bibinfo {volume} {56}} (\bibinfo
  {year} {2020}),\ e2019WR026279}\BibitemShut {NoStop}%
\end{thebibliography}%
\end{document}